\documentclass[aps,pra,twocolumn,showpacs,groupedaddress]{revtex4}
\usepackage{graphicx}
\usepackage{amssymb}
\usepackage{mathrsfs}
\usepackage{amsmath}
\usepackage{bm,hyperref}

\begin{document}
\title{Berry Phase and Hannay's Angle in a Quantum-Classical Hybrid System}
\author{H. D. Liu, S. L. Wu and X. X. Yi\footnote{yixx@dlut.edu.cn}}
\affiliation{School of Physics and Optoelectronic Technology,
              Dalian University of Technology, Dalian 116024, China}

\date{\today}

\begin{abstract}
Berry phase, which had been discovered for more than two decades,
provides us a very deep insight on the geometric structure of
quantum mechanics. Its classical counterpart--Hannay's angle
 is defined if closed curves of  action
variables return to the same curves in phase space after a time
evolution. In this paper, we study the Berry phase and Hannay's
angle in a quantum-classical hybrid system under the
Born-Oppenheimer approximation. By quantum-classical hybrid system,
we denote a composite system consists of a quantum subsystem and a
classical subsystem. The effects of subsystem-subsystem couplings on
the Berry phase and Hannay's angle are explored. The results show
that the Berry phase has been changed sharply by the couplings,
whereas the couplings have small effect on the Hannay's angle.
\end{abstract}
 \maketitle

\section{Introduction}
The concept of Berry phase has paved a new way for understanding
quantum physics, it has attracted much attention since the Berry's
discovery \cite{Chruscinski,Berry,Berry1}  and found potential
applications in fields ranging from chemistry to condensed matter
physics. Shortly after Berry's discovery, Simon gave  a mathematical
interpretation to this phase that it can be regarded as the holonomy
in a Hermitian line bundle since the adiabatic theorem naturally
defines a connection in such a bundle\cite{Simon}.  After these
remarkable discoveries, Hannay found that this geometrical phase not
only exist in quantum system but also  in classical world
\cite{Hannay}. Analogous with the Berry phase, the angle variable of
classical integrable systems \cite{Arnold} acquires an additional
angle shift as the system slowly cycles in phase space. This angle
shift is called Hannay's angle. It was later proved by Berry that
the geometric phase and Hannay's angle possess a natural relation
under semiclassical approximation \cite{Berry2}. As a matter of
course, this quantum-classical correspondence gives rise to many
impressive explorations\cite{Giavarini,Jarzynski,Pati}.

It is known that a quantum  Hilbert space carries the same structure
of a classical phase space \cite{Chruscinski,Arnold}. By this
virtue, a quantum system can be treated like a classical system
without loss of physics \cite{Heslot,Weinberg}. Particularly in Ref.
\cite{Weinberg}, the authors  introduced a general framework for
testing nonlinear quantum mechanics. In this generalized theory, the
elements of quantum mechanics, such as wave functions, observable,
symmetries and time evolution, all can be treated classically. Based
on this instructive work, some interesting efforts have been devoted
to nonlinear quantum system  and quantum-classical hybrid system
\cite{Polchinski,Wu,Zhang}. Here come the questions: Since quantum
mechanics can be put into the framework of classic mechanics, can
Berry phase be presented into the form of Hannay's angle? What is
the Berry phase or Hannay's angle of the quantum-classical hybrid
system? How does the subsystem-subsystem couplings affect the
Hannay's angle and the Berry phase? We shall shed light on these
questions in this paper.

The paper is organized as follows. In Sec. {\rm II}, we first
represent a quantum system in the framework of classical theory by
Weinberg's method\cite{Weinberg}, then we calculate the Hannay's
angle of the  quantum system. Next we compare this Hannay's angle
with the Berry phase of the original quantum system and find that
the Hannay's angle and Berry phase  differ only in a sign\cite{Wu}.
An example is given to show this result.  In Sec. {\rm III}, we
represent a quantum-classical  hybrid system in classical mechanics
based on the Born-Oppenheimer approximation, a unified one-form
which can deduce both of the Berry phase and the Hannay's angle  is
given. By this one-form, we calculated the Hannay's angles and Berry
phase and study the effect of subsystem-subsystem couplings on the
Hannay's angle. Finally, we conclude our results  In Sec. {\rm IV}.

\section{Berry phase and Hannay's angle}
In quantum theory, observables   are represented by Hermitian
matrices $F$ or real bilinear functions $\langle \psi|\hat
F|\psi\rangle$. Weinberg has generalized the representation  into
real non-bilinear functions $f(\bm{\psi},\bm{\psi}^*)$ to include
nonlinearity. Let $\{|\varphi_n\rangle\}$ be an orthonormal basis of
Hilbert space, the  Schr\"{o}dinger equation can be rewritten as:
\begin{equation}
i\hbar\frac{d\psi_n}{dt}=\frac{\partial h}{\partial\psi_n^*},
\label{seq}
\end{equation}
with $|\psi\rangle=\sum\psi_n|\varphi_n\rangle$,
$\bm{\psi}=(\psi_1,\ldots,\psi_n,\ldots,\psi_N)^T$ and
$h(\bm{\psi},\bm{\psi}^*)$ is the total energy of the system. If we
decompose $\psi_n$ into real and imaginary parts
$\psi_n=(q_n+ip_n)/\sqrt{2\hbar}$, the schr\"{o}dinger equation and
its complex conjugate can be written as Hamiltonian  canonical
equations \cite{Heslot,Weinberg}
\begin{equation}
\dot{q}_n=\frac{\partial h}{\partial p_n},~\dot{p}_n=-\frac{\partial
h}{\partial q_n}.
\end{equation}
The Hamiltonian function $h(\psi,\psi^*)$ can be transformed into
$h(\bm{q},\bm{p})$. It is amazing to note that the quantum hermitian
structure becomes the symplectic structure of classical mechanics,
$dp_n\wedge dq_n=i\hbar d\psi^*_n\wedge d\psi_n$.

Now, let us  turn to the first question. Consider a quantum system
with $N$ levels, whose Hamiltonian function $h(\psi,\psi^*,\bm{X})$
depends on a set of slowly varying parameters $\bm{X}=(X_1,
Y_1,\ldots)$ and its quantum state $|\psi\rangle$. By the procedure
mentioned above, i.e., by decomposing  $|\psi\rangle$ with  basis
$\{|n\rangle\}$, $|\psi\rangle=\sum_n \psi_n(t)|n\rangle$, and
setting $\psi_n(t)=(q_n(t)+ip_n(t))/\sqrt{2\hbar}$, we can write the
Hamiltonian function in terms  of "position variable $\bm{q}$" and
"momentum variable $\bm{p}$" as $h(\bm{p}(t),\bm{q}(t))$. Of course,
the state vector $|\psi\rangle$ can also be expanded taking  the
Hamiltonian eigenstates $|E_k(\bm{X})\rangle$ as a basis,
$|\psi\rangle=\sum_k\psi_k'(t)|E_k(\bm{X})\rangle$. By the adiabatic
theorem, the occupation probability of each eigenstate
$|\psi_k'(t)|^2$ remains unchanged in the adiabatic  limit. One then
can introduce a new pair of variables $(\bm{\theta},\bm{I})$ by
\begin{equation}
\psi_k'(t)=\sqrt{\frac{I_k}{\hbar}}e^{-i\theta_k},\label{itheta}
\end{equation}
and write the Hamiltonian function as \cite{Weinberg,Wu}
\begin{equation}
\bar{h}=\bar{h}(\bm{I},\bm{X})=\sum_kE_k(\bm{X})I_k/\hbar,
\end{equation}
where $E_k(\bm{X})$ is the eigenvalue of the Hamiltonian $\hat H$
with corresponding  eigenstate $|E_k(\bm{X})\rangle$. It has been
proved that the two new variables $\bm{\theta}$ and $\bm{I}$ satisfy
the same canonical equations as the  angle-action variables in
classical mechanics\cite{Weinberg},
\begin{equation}
\dot{\theta}_k=\frac{\partial \bar{h}}{\partial I_k},~\dot{I}_k=0.
\end{equation}
So far, the quantum Hamiltonian $\hat H$ was transformed into
classical Hamiltonian function $\bar{h}(\bm{I},\bm{X})$, and the
quantum unitary transformation $|E_k(\bm{X})\rangle=\sum_n
C_{kn}(\bm{X})|n\rangle$ turns out to be a classical canonical
transformation $(\bm{q},\bm{p})\rightarrow (\bm{\theta},\bm{I})$.
\begin{equation}
\begin{aligned}
&p_n=\sum_k\sqrt{2I_k}\left[\cos\theta_k
\mathrm{Im}(C_{kn}(\bm{X}))-\sin\theta_k
\mathrm{Re}(C_{kn}(\bm{X}))\right],\\
&q_n=\sum_k\sqrt{2I_k}\left[\cos\theta_k
\mathrm{Re}(C_{kn}(\bm{X}))+\sin\theta_k
\mathrm{Im}(C_{kn}(\bm{X}))\right]. \label{canonical}
\end{aligned}
\end{equation}
According to Berry's theory \cite{Berry2}, the Hannay's angle of the
system can be written as
\begin{equation}
\Delta\theta_k(\bm{I};\bm{X})=-\frac{\partial}{\partial I_k}\oint
A_H(\bm{I};\bm{X}),
\end{equation}
where
\begin{equation}
\begin{aligned}
A_H(\bm{I};\bm{X})&=\langle
\bm{p}(\bm{\theta},\bm{I};\bm{X})d_{\bm{X}}\bm{q}
(\bm{\theta},\bm{I};\bm{X})\rangle_\theta \\
&=\frac 1{(2\pi)^N}\oint d\theta
\sum_np_n(\bm{\theta},\bm{I};\bm{X})d_{\bm{X}}q_n
(\bm{\theta},\bm{I};\bm{X})\label{Hangle}
\end{aligned}
\end{equation}
is the angle  one-form for Hannay's angle \cite{Gozzi}. The angular
brackets $\langle\cdots\rangle_\theta$ denote an averaging over all
angles $\bm{\theta}$, and $d_{\bm{X}}$ is defined as
$d_{\bm{X}}F(\bm{X})=\frac{\partial F(\bm{X})}{\partial \bm{X}}\cdot
d{\bm{X}}$. Substituting Eq. (\ref{canonical}) into Eq.
(\ref{Hangle}), we obtain
\begin{equation}
\begin{aligned}
A_H&=\sum_k I_k\sum_n\left[i(C_{kn}^*(\bm{X})d_{\bm{X}}C_{kn}(\bm{X}))\right]\\
&=\sum_k iI_k \langle
E_k(\bm{X})|d_{\bm{X}}E_k(\bm{X})\rangle=\sum_kiI_kA_B(k;\bm{X}).
\end{aligned}\label{aia}
\end{equation}
We note that $A_B(k;\bm{X})$ is nothing but the one-form for Berry
phase\cite{Gozzi}, this  means that the Hannay's angles exactly
equal to the minus  Berry phases of the original quantum system
\cite{Wu},
\begin{equation}
\Delta\theta_k(\bm{I};\bm{X})=-\frac\partial{\partial I_k}\oint
A_H=-i\oint A_B(k;\bm{X})=-\gamma_k(C).\label{anglephase}
\end{equation}
 The appearance of the
minus  is because the angle variables of the effective classical
system  correspond to the opposite numbers of the phases in Eq.
(\ref{itheta}). Therefore, we can choose $A(\bm{I};\bm{X})=A_H$ to
be a general one-form that
\begin{equation}
\Delta\theta_k=-\gamma_k=-\frac\partial{\partial I_k}\oint
A(\bm{I};\bm{X}).\label{gena}
\end{equation}
To shed more light on the result, we consider the adiabatic
evolution of a spin-half particle with  magnetic moment $\mu$ in an
external magnetic field $\bm{B}$. The Hamiltonian reads
\begin{equation}
\hat H=-\mu\hat{\bm{\sigma}}\cdot\bm{B},\label{spinhq}
\end{equation}
where $\hat{\bm{\sigma}}=(\hat\sigma_1,\hat\sigma_2,\hat\sigma_3)$
are Pauli Matrices. As aforementioned,  if we choose the two spin
eigenstates $|\pm\rangle$ as the basis, the Hamiltonian in Eq.
(\ref{spinhq}) can be transformed into a Hamiltonian function
\begin{equation}
\begin{aligned}
h(\bm{p},\bm{q};\bm{B})=&\frac1\hbar[(q_1q_2+p_1p_2)B_1+(p_1q_2-p_2q_1)B_2\\
&+\frac12(p^2_2+q^2_2-p^2_1-q^2_1)B_3],
\end{aligned}
\end{equation}
with $|\psi\rangle=\psi_1|-\rangle+\psi_2|+\rangle$ and $
\psi_j=(q_j+ip_j)/\sqrt{2\hbar}$, ($j=1,2$). The canonical variables
$(\bm{q},\bm{p})$ satisfy the normalization condition \cite{Heslot},
\begin{equation}
\sum_{j=1}^2(p^2_j+q^2_j)=2\hbar.
\end{equation}
It is interesting to note that by defining  a vector,
$\bm{S}=(S_1,S_2,S_3)$, the Hamiltonian function can be written as,
\begin{equation}
h(\bm{S};\bm{B})=-\mu\bm{S}\cdot\mathbf{B},
\end{equation}
where
\begin{equation}
\left\{
  \begin{array}{l}
    S_1=(q_1q_2+p_1p_2)/\hbar, \\
    S_2=(p_1q_2-p_2q_1)/\hbar, \\
    S_3=\left(p^2_2+q^2_2-p^2_1-q^2_1\right)/(2\hbar). \\
  \end{array}
\right.
\end{equation}
The normalization condition in terms of $\bm{S}$ is  $S^2\equiv
S^2_1+S^2_2+S^2_3=1$, and their Poisson bracket has a relation with
the quantum commutator as \cite{Heslot}
\begin{equation}
\{S_i,S_j\}=2\epsilon_{ijk}S_k/\hbar=\frac1{i\hbar}\langle\psi|[\hat\sigma_i,\hat\sigma_j]|\psi\rangle.
\end{equation}
Moreover, if we choose $|\pm\rangle$ as the basis, $\bm{S}$ is
nothing but the Stokes parameters which span the Poincare sphere,
\begin{equation}
\left\{
  \begin{array}{l}
    I=|\psi_2|^2+|\psi_1|^2=S^2=1, \\
    U=2\mathrm{Re}(\psi_2\psi_1^*)=S_1S,\\
    V=2\mathrm{Im}(\psi_2\psi_1^*)=S_2S,\\
    Q=|\psi_2|^2-|\psi_1|^2=S_3S.\\
  \end{array}
\right.
\end{equation}

We now move to calculate the Hannay's angle. Since the Hamiltonian
in Eq. (\ref{spinhq}) has two eingenstates
\begin{eqnarray}
&&|E_1\rangle=\sqrt{\frac{B+B_3}{2B}}|+\rangle+\frac{B_1+iB_2}{\sqrt{2B(B+B_3)}}|-\rangle,\nonumber\\
&&|E_2\rangle=-\sqrt{\frac{B-B_3}{2B}}|+\rangle+\frac{B_1+iB_2}{\sqrt{2B(B-B_3)}}|-\rangle,\label{spineigen}
\end{eqnarray}
with eigenenergies $-\mu B$ and $\mu B$, respectively, where
$B=\sqrt{B_1^2+B^2_2+B_3^2}$. The canonical transformation
$(\bm{q},\bm{p})\rightarrow (\bm{\theta},\bm{I})$ and the
transformed Hamiltonian function can be written as
\begin{equation}
\begin{aligned}
q_1=&\sqrt{2I_1}\left[\frac{B_1\cos\theta_1}{\sqrt{2B(B+B_3)}}+\frac{B_2\sin\theta_1}{\sqrt{2B(B+B_3)}}\right]\\
&+\sqrt{2I_2}\left[\frac{B_1\cos\theta_2}{\sqrt{2B(B-B_3)}}+\frac{B_2\sin\theta_2}{\sqrt{2B(B-B_3)}}\right], \\
q_2=&\sqrt{\frac{2I_1(B+B_3)}{2B}}\cos\theta_1-\sqrt{\frac{2I_2(B-B_3)}{2B}}\cos\theta_2,\\
p_1=&\sqrt{2I_1}\left[\frac{B_2\cos\theta_1}{\sqrt{2B(B+B_3)}}-\frac{B_1\sin\theta_1}{\sqrt{2B(B+B_3)}}\right]\\
&+\sqrt{2I_2}\left[\frac{B_2\cos\theta_2}{\sqrt{2B(B-B_3)}}-\frac{B_1\sin\theta_2}{\sqrt{2B(B-B_3)}}\right],\\
p_2=&\sqrt{\frac{2I_2(B-B_3)}{2B}}\sin\theta_2-\sqrt{\frac{2I_1(B+B_3)}{2B}}\sin\theta_1,\\
\bar{h}(\bm{I};\bm{B})&=\mu B(I_2-I_1),\\
\end{aligned}\label{spincanon}
\end{equation}
with $\bm{q}=(q_1,q_2)$, $\bm{p}=(p_1,p_2)$,
$\bm{\theta}=(\theta_1,\theta_2)$ and $\bm{I}=(I_1,I_2)$. Therefore,
we obtain the angle one-form by Eq. (\ref{Hangle}),
\begin{equation}
A=\frac{B_2dB_1-B_1dB_2}{2B(B+B_3)}I_1+\frac{B_2dB_1-B_1dB_2}{2B(B-B_3)}I_2.
\end{equation}
The Hannay's angles can thus be obtained by Eq. (\ref{gena})
\begin{eqnarray}
&&\Delta\theta_1=-\oint\frac{B_2dB_1-B_1dB_2}{2B(B+B_3)},\nonumber\\
&&\Delta\theta_2=-\oint\frac{B_2dB_1-B_1dB_2}{2B(B-B_3)},
\end{eqnarray}
which differ from the Berry phases for the original quantum
Hamiltonian \cite{Chruscinski} only by a sign.

\section{Berry phase and Hannay's angle in hybrid system}
Based on formalism in the last section, we now turn to the second
question risen in the Introduction. Consider a hybrid system
consisting of a classical and quantum subsystem,  the Hamilton
function of this  quantum-classical hybrid system under
Born-Oppenheimer approximation can be written as \cite{Zhang,Zhan}
\begin{equation}
H_{hybrid}=\langle\psi|\hat
H_1(\bm{Q},\bm{X}_1)|\psi\rangle+H_2(\bm{P},\bm{Q};\bm{X}_2),
\end{equation}
where $|\psi\rangle$ is the  state of the fast quantum subsystem.
The Hamiltonian function $H_2$ describes a slow classical subsystem
with momentum $\bm{P}$ and coordinate $\bm{Q}$, $\bm{X}_1=(X_1,
Y_1,\ldots)$ and $\bm{X}_2=(X_2, Y_2\ldots)$ are slowly varying
parameters of the quantum and classical subsystems, respectively.
The subsystem-subsystem coupling is included in $\hat
H_1(\bm{Q},\bm{X}_1)$.  Following the procedure given in the last
section, we
 first choose a  basis, then expand the state $|\psi\rangle$
in this basis, next decompose the expansion coefficients into real
parts $\bm{q}$ and imaginary parts $\bm{p}$, finally represent the
hybrid system by a classical Hamiltonian function,
\begin{equation}
H=H_1(\bm{p},\bm{q};\bm{Q},\bm{X}_1)+H_2(\bm{P},\bm{Q};\bm{X}_2).
\end{equation}
As known, the Hamiltonian of quantum subsystem
$H_1(\bm{p},\bm{q};\bm{Q},\bm{X}_1)$ can be transformed into
$\bar{H}_1(\bm{\theta},\bm{I};\bm{Q},\bm{X}_1)$ by a canonical
transformation $(\bm{q},\bm{p})\rightarrow(\bm{\theta},\bm{I})$. But
the new Hamiltonian $\bar{H}_1$ differs from the old one, it takes
\cite{Berry2,Arnold},
\begin{equation}
\begin{aligned}
\bar{H}_1(\bm{\theta},\bm{I};\bm{Q},\bm{X}_1)=&\mathscr{H}_1(\bm{I};\bm{Q},\bm{X}_1)+\frac{\partial
S}{\partial t}\\
=&\mathscr{H}_1(\bm{I};\bm{Q},\bm{X}_1)+\dot{\bm{Q}}\cdot\left(\frac{\partial\mathscr{S}}{\partial
\bm{Q}}-\bm{p}\frac{\partial \bm{q}}{\partial
\bm{Q}}\right)\\
&+\dot{\bm{X}}_{1}\cdot\left(\frac{\partial\mathscr{S}}{\partial
\bm{X}_{1}}-\bm{p}\frac{\partial \bm{q}}{\partial
\bm{X}_{1}}\right), \label{Htrans}
\end{aligned}
\end{equation}
where
\begin{equation}
\mathscr{H}_1(\bm{I};\bm{Q},\bm{X}_1)\equiv
H_1(\bm{p}(\bm{\theta},\bm{I};\bm{Q},\bm{X}_1),\bm{q}(\bm{\theta},\bm{I};\bm
{Q},\bm{X}_1);\bm{Q},\bm{X}_1),
\end{equation}
and $S(\bm{q},\bm{I};\bm{Q},\bm{X}_1)$ is the generating function of
the transformation and the single-valued function $\mathscr{S}\equiv
S(\bm{q}(\bm{\theta},\bm{I};\bm{Q},\bm{X}_1),\bm{I};\bm{Q},\bm{X}_1)$
is introduced to give an explicit form for $\bar{H}_1$\cite{Berry2}.
Since the variables $\bm{Q}$ can be treated as slowly varying
parameters like $\bm{X}_1$, an average  over $\theta$ may be taken
to approximate to Eq. (\ref{Htrans}) \cite{Arnold},
\begin{equation}
\langle
\bar{H}_1\rangle_\theta=\mathscr{H}_1(\bm{I};\bm{Q},\bm{X}_1)
-\dot{\bm{Q}}\cdot\left\langle\bm{p}\frac{\partial \bm{q}}{\partial
\bm{Q}}\right\rangle_\theta
-\dot{\bm{X}}_{1}\cdot\left\langle\bm{p}\frac{\partial
\bm{q}}{\partial \bm{X}_{1}}\right\rangle_\theta, \label{Hav}
\end{equation}
where the angular brackets $\langle\cdots\rangle_\theta$ denote an
averaging over all angles $\bm{\theta}$, and the terms
$\dot{\bm{Q}}\cdot\langle\frac{\partial\mathscr{S}}{\partial
\bm{Q}}\rangle_\theta$ and
$\dot{\bm{X}}_1\cdot\langle\frac{\partial\mathscr{S}}{\partial
\bm{X}_1}\rangle_\theta$ are dropped for zero contribution to the
equations of motion. Thus, the Hamiltonian of total system can be
written as
\begin{equation}
\begin{aligned}
H_{av}=&\mathscr{H}_1(\bm{I};\bm{Q},\bm{X}_1)+H_2(\bm{P},\bm{Q};\bm{X}_2)\\
&-\dot{\bm{Q}}\cdot\left\langle \sum_ip_i\frac{\partial
q_i}{\partial \bm{Q}}\right\rangle-\dot{\bm{X}}_{1}\cdot\left\langle
\sum_ip_i\frac{\partial q_i}{\partial \bm{X}_{1}}\right\rangle.
\end{aligned}\label{cla}
\end{equation}
where the first two terms is the effective Hamiltonian under
Born-Oppenheimer approximation and the latter two terms are related
to  the Berry phase of the quantum subsystem, because the quantum
one-form is defined as
$A_1(\bm{I};\bm{X_1},Q)=-\left(\dot{\bm{Q}}\cdot\left\langle
\sum_ip_i\frac{\partial q_i}{\partial
\bm{Q}}\right\rangle+\dot{\bm{X}}_{1}\cdot\left\langle
\sum_ip_i\frac{\partial q_i}{\partial
\bm{X}_{1}}\right\rangle\right) dt$ like Eq. (\ref{gena}).

Since $\bm{I}$ can be treated as a constant in the adiabatic limit,
the Hamiltonian in Eq. (\ref{cla}) only contains the variables of
the classical subsystem. Therefore, we can transform this
Hamiltonian into a function of angle-action variables via the
canonical transformation
$(\bm{Q},\bm{P})\rightarrow(\bm{\phi},\bm{J})$. By averaging over
$\bm{\phi}$, we finally obtain an averaged Hamiltonian for the
hybrid system,
\begin{equation}
\langle H_{av}\rangle=\mathscr{H}(\bm{I},\bm{J};\bm{X}_1,\bm{X}_2)
+\frac{A(\bm{I},\bm{J};\bm{X}_1,\bm{X}_2)}{dt},
\end{equation}
with $\mathscr{H}(\bm{I},\bm{J};\bm{X}_1,\bm{X}_2)\equiv
\mathscr{H}_1(\bm{I};\bm{Q}(\bm{\phi},\bm{J};\bm{X}_1,\bm{X}_2),
\bm{X}_1)+H_2(\bm{Q}(\bm{\phi},\bm{J};\bm{X}_1,\bm{X}_2),
\bm{P}(\bm{\phi},\bm{J};\bm{X}_1,\bm{X}_2);\bm{X}_2)$. According to
 classical mechanics \cite{Arnold}, the time-dependent canonical
transformation will bring out a geometric term in the classical
Hamiltonian. This together with the quantum one-form $A_1$, defines
a general one-form for the whole system,
\begin{equation}
A(\bm{I},\bm{J};\bm{X}_1,\bm{X}_2)\equiv\left\langle
A_1\right\rangle_\phi-\dot{\bm{X}}_{2}\cdot\left\langle
\sum_iP_i\frac{\partial Q_i}{\partial \bm{X}_{2}}\right\rangle_\phi
dt.
\end{equation}
By this one-form, we can obtain   the Berry phases $\gamma_k$ and
the Hanny's angles $\Delta\phi_l$ uniformly for the hybrid system by
Eq. (\ref{anglephase})
\begin{equation}
\gamma_k=\frac\partial{\partial I_k}\oint
A,~~~\Delta\phi_l=-\frac\partial{\partial J_l}\oint A.\label{hybrid
angle}
\end{equation}

To illustrate the result in Eq. (\ref{hybrid angle}), we now
consider a spin-half particle in an external magnetic field $\bm{B}$
coupled with a classical harmonic oscillator. This model is widely
used since  the begining of quantum mechanics (e.g., Ref
\cite{Leggett}). The Hamiltonian for such a system reads
\begin{equation}
H=\langle\psi|\hat{H}_1|\psi\rangle +(XQ^2+2YPQ+ZP^2),\label{so}
\end{equation}
where
$\hat{H}_1=-\mu\hat{\bm{\sigma}}\cdot\bm{B}-\mu\lambda\hat{\sigma}_3Q$
is the Hamiltonian of the quantum particle coupled to the magnetic
field $\bm{B}\equiv B(\cos\varphi,\sin\varphi,0)$ with magnetic
moment $\mu$ and coupling constant $\lambda$, and $\bm{X}=(X, Y,Z)$
are the time-dependent parameters of the classical subsystems. The
quantum state $|\psi\rangle$ is defined by the spin eigenstates
$|\pm\rangle$. I.e.,
$|\psi\rangle=[(q_-+ip_-)|-\rangle+(q_++ip_+)|+\rangle]/\sqrt{2\hbar}
$. It is easy to get  the eigenfunctions $|E_\pm\rangle$ and the
corresponding eigenvalues for $\hat{H}_1$,
\begin{equation}
\begin{aligned}
&E_{\pm}=\pm\mu B_{tot},\\
&|E_+\rangle=\cos\frac{\Theta}{2}|+\rangle
+e^{i\varphi}\sin\frac{\Theta}{2}|-\rangle,\\
&|E_-\rangle=-\sin\frac{\Theta}{2}|+\rangle
+e^{i\varphi}\cos\frac{\Theta}{2}|-\rangle,
\end{aligned}
\end{equation}
with $B_{tot}\equiv\sqrt{B^2+\lambda^2Q^2}$ and
$\cos\Theta\equiv\lambda Q/B_{tot}$. Following the proposed
procedures, we transform the quantum Hamiltonian $\hat{H}_1$ into a
classical form with a canonical transformation
$(q_\pm,p_\pm)\rightarrow(\theta_\pm,I_\pm)$, where
$|\psi\rangle=\sqrt{I_+/\hbar}e^{-i\theta_+}|E_-\rangle+\sqrt{I_-/\hbar}e^{-i\theta_-}|E_+\rangle
$. After averaging over the angles $\theta_\pm$, we obtain
\begin{equation}
\begin{aligned}
H_{av}=&\mu B_{tot}(I_+-I_-)+\frac{A_1(\bm{I};Q,\bm{B})}{dt}\\
&+\frac12(XQ^2+2YPQ+ZP^2),
\end{aligned}\label{soav}
\end{equation}
where
\begin{equation}
A_1(\bm{I};Q,\bm{B})=-\frac{1}{2}\left[I_+(1-\cos\Theta)+I_-(1+\cos\Theta)\right]d\varphi
\end{equation}
is the phase one-form for the quantum subsystem. So far, the
Hamiltonian is fully classical and only contains  variables of
classical subsystem. It is difficult to derive the action variable
for this Hamiltonian, because the dependence of $B_{tot}$ on $Q$ is
complicated. However,  in the weak coupling limit $\lambda\ll |\frac
BQ|$, the problem becomes easy. Expanding $B_{tot}$  to the second
order in  $\lambda$,
\begin{equation}
B_{tot}\approx B+\frac{\lambda^2Q^2}{2B},
\end{equation}
we obtain the Hamiltonian in the weak coupling limit,
\begin{equation}
\begin{aligned}
H_{av}\approx&\mu B(I_+-I_-)+\frac{\mu\lambda^2Q^2(I_+-I_-)}{2B}+\frac{A_1(\bm{I};Q,\bm{B})}{dt}\\
&+\frac12(XQ^2+2YPQ+ZP^2).
\end{aligned}
\end{equation}
According to Berry's theory\cite{Berry2}, we introduce the canonical
transformation $(Q,P)\rightarrow(\phi,J)$
\begin{equation}
\begin{aligned}
&Q=\left(\frac{2ZJ}{\Omega}\right)^{1/2}\cos\phi,\\
&P=-\left(\frac{2ZJ}{\Omega}\right)^{1/2}\left(\frac
YZ\cos\phi+\frac\Omega Z\sin\phi\right),\label{sojphi}
\end{aligned}
\end{equation}
with $\Omega\equiv\{[X+\mu\lambda^2(I_+-I_-)/B]Z-Y^2\}^{1/2}$.
Substituting Eq. (\ref{sojphi}) into Eq. (\ref{soav}) and averaging
it over $\phi$, we obtain the Hamiltonian for the hybrid system,
\begin{equation}
\langle H_{av}\rangle=\mu
B(I_+-I_-)+J\Omega+\frac{A(\bm{I},J;\bm{B},\bm{X})}{dt},
\end{equation}
where the general one-form for the hybrid system is given by
($\langle\cos\Theta\rangle_\phi=0$),
\begin{equation}
\begin{aligned}
A(\bm{I},J;\bm{B},\bm{X})=-\frac{(I_++I_-)}{2}d\varphi
-\frac{YJ}{2Z}d\left(\frac{Z}{\Omega}\right).
\end{aligned}
\end{equation}
Thus the Berry phases and Hannay's angle can be given uniformly by a
straightforward calculation
\begin{eqnarray}
&&\gamma_\pm=\oint\left[-\frac12d\varphi\mp\frac{\mu\lambda^2
Z^2J}{4\Omega^3B}d\left(\frac{Y}{Z}\right)\right],\nonumber\\
&&\Delta\phi=\oint\frac{Y}{2Z}d\left(\frac{Z}{\Omega}\right).
\end{eqnarray}
The range of integration is determined  by the common period of
$\bm{X}$ and $\bm{B}$. It is easy to find that when $\lambda=0$, the
Berry phases are the solid angle $\pi$ on Bloch sphere. The
subsystem-subsystem couplings not only adds a correction to the
Berry phases,  but also change range of integration. Whereas the
interaction gives the Hannay's angle a modification
$\omega\equiv(XZ-Y^2)^{1/2}\rightarrow\Omega$ through $\Omega$.

The second example is a quantum harmonic oscillator coupled with a
classical one. The mean-field Hamiltonian \cite{Zhan} for this
hybrid system is
\begin{equation}
\hat{H}=\langle\psi|\hat{H}_1|\psi\rangle+\frac12(X_2Q^2+2Y_2PQ+Z_2P^2),
\end{equation}
where
$\hat{H}_1=\frac12[X_1\hat{q}^2+Y_1(\hat{p}\hat{q}+\hat{q}\hat{p})+Z_1\hat{p}^2]+K\hat{q}Q$
includes  the free Hamiltonian of the quantum subsystem and the
subsystem-subsystem coupling, $K$ is the coupling constant and
$\bm{X}_1=(X_1, Y_1,Z_1)$, $\bm{X}_2=(X_2,Y_2,Z_2)$ are the
time-dependent parameters of the two subsystems. Similar with Ref.
\cite{Berry2}, the eigenfunctions and eigenvalues of $\hat{H}_1$ are
\begin{eqnarray}
E_n&=&\left(n+\frac12\right)\hbar\omega-\frac{Z_1K^2Q^2}{2\omega^2},\nonumber\\
\psi_n&=&\sqrt{\alpha}\chi_n\left(\alpha\left(q+\frac{KZ_1Q}{\omega^2}\right)\right)
\exp\left(\frac{-iY_1q^2}{2Z_1\hbar}\right),\nonumber\\
\label{qcoeigen}
\end{eqnarray}
with $\omega=\sqrt{X_1Z_1-Y_1^2}$,
$\alpha=\sqrt{\frac{\omega}{Z_1\hbar}}$ and the normalized Hermite
functions $\chi_n(\xi)$. By the transformation,
$(\bm{q},\bm{p})\rightarrow(\bm{\theta},\bm{I})$, we obtain the
averaged Hamiltonian by averaging over the angles $\bm{\theta}$,
\begin{equation}
\begin{aligned}
H_{av}=&\sum_n(n+1/2)
I_{n}\omega-\frac{Z_1K^2Q^2}{2\omega^2}+\frac{A_1(\bm{I};\bm{X_1},Q)}{dt}\\
&+\frac12(X_2Q^2+2Y_2PQ+Z_2P^2),\label{exhav}
\end{aligned}
\end{equation}
where
\begin{equation}
\begin{aligned}
A_1&(\bm{I};\bm{X}_1,Q)\\
&=\sum_nI_{n}\left[\frac{(2n+1)Z_1}{4\omega}+\frac{K^2Z_1^2Q^2}{2\hbar
\omega^4}\right]d\left(\frac{Y_1}{Z_1}\right)
\end{aligned}
\end{equation}
is the phase one-form for the quantum subsystem. Note that the
action variables $\bm{I}$ are invariant, we can treat them as
constants. After the canonical transformation
$(Q,P)\rightarrow(\phi,J)$ (elliptic case)
\begin{equation}
\begin{aligned}
Q&=\left(\frac{2Z_2J}{\Omega}\right)^{1/2}\cos\phi,\\
P&=-\left(\frac{2Z_2J}
{\Omega}\right)^{1/2}\left(\frac{Y_2}{Z_2}\cos\phi+\frac{\Omega}{Z_2}\sin\phi\right),
\end{aligned}\label{QPcanon}
\end{equation}
and substituting Eq. (\ref{QPcanon}) into  Eq. (\ref{exhav}) as well
as  averaging over $\bm{\phi}$, the Hamiltonian of the hybrid system
becomes
\begin{equation}
\langle H_{av}\rangle=\sum_n\left(n+\frac12\right)
I_{n}\omega+J\Omega+\frac{A(\bm{I},J;\bm{X_1},\bm{X_2})}{dt}
\end{equation}
with
$\Omega=[\frac{(\omega^2X_2-Z_1K^2)Z_2}{\omega^2}-Y_2^2]^{1/2}$, and
the general one-form for the hybrid system is
\begin{equation}
\begin{aligned}
A(\bm{I},J;\bm{X_1},\bm{X_2})=&\sum_nI_{n}\left[\frac{(2n+1)Z_1}{4\omega}+\frac{K^2Z_1^2Z_2J}{2\hbar
\omega^4\Omega}\right]\\
&\cdot
d\left(\frac{Y_1}{Z_1}\right)-\frac{Y_2J}{2Z_2}d\left(\frac{Z_2}{\Omega}\right).
\end{aligned}
\end{equation}
Therefore the Berry phases and Hannay's angle can be given by
\begin{eqnarray}
&&\gamma_n=\oint\left[\frac{(2n+1)Z_1}{4\omega}+\frac{K^2Z_1^2Z_2J}{2\hbar
\omega^4\Omega}\right]d\left(\frac{Y_1}{Z_1}\right),\nonumber\\
&&\Delta\phi=\oint\left[\frac{Y_2}{2Z_2}d\left(\frac{Z_2}{\Omega}\right)-\frac{K^2Z_1^2Z_2}{2
\omega^4\Omega}d\left(\frac{Y_1}{Z_1}\right)\right]. \label{pa}
\end{eqnarray}
The limits of the integrals are decided by the common period of
$\bm{X}_1$ and $\bm{X}_2$.

\begin{figure}
\includegraphics*[width=0.95\columnwidth,
height=0.6\columnwidth]{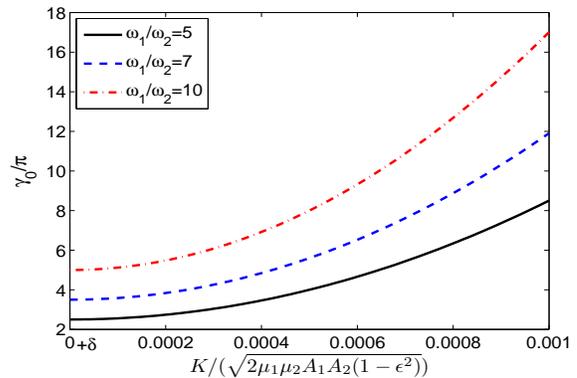} \caption{(color online) The Berry
phase of the ground state $\gamma_0$ as a function of coupling
constant $K$. The parameters are $\epsilon=\sqrt3/2$,
$A_1/A_2=10^{8}$, $J/\hbar=10^{13}$, and $\omega_1/\omega_2$ takes
different values for different lines.}\label{gamma}
\end{figure}

\begin{figure}
\includegraphics*[width=0.95\columnwidth,
height=0.6\columnwidth]{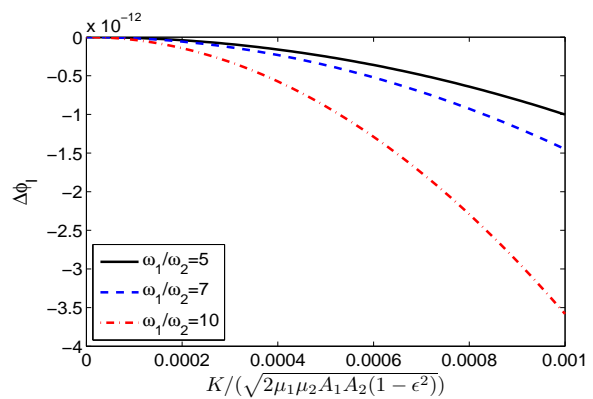} \caption{(color online) The
coupling Hannay's angle $\Delta\phi_I$ as a function of coupling
constant $K$, with the same parameters as
Fig.\ref{gamma}.}\label{theta}
\end{figure}

Now we take a specific choice of the periodic parameters to see an
exact result of our theory. Set
\begin{eqnarray}
\left\{
  \begin{array}{l}
    X_1=A_1\mu_1(1+\epsilon\cos(\omega_1t)) \\
    Y_1=-A_1\epsilon\sin(\omega_1t) \\
    Z_1=\frac{A_1}{\mu_1}(1-\epsilon\cos(\omega_1 t))\\
  \end{array}
\right., \nonumber\\
\left\{
\begin{array}{l}
    X_2=A_2\mu_2(1+\epsilon\cos(\omega_2t)) \\
    Y_2=-A_2\epsilon\sin(\omega_2t) \\
    Z_2=\frac{A_2}{\mu_2}(1-\epsilon\cos(\omega_2t)) \\
  \end{array}
\right.,
\end{eqnarray}
where $\omega_1$, $\omega_2$ are the frequencies of the parameters,
$\epsilon$ is a dimensionless constant and the units of $A$ and
$\mu$ are $s^{-1}$ and $kg/s$, respectively. If the frequency ratio
$\omega_1/\omega_2$ is rational, $\bm{X}_1$ and $\bm{X}_2$ have a
common period $T$. After a straightforward calculation, we obtain
the phases and angle in Eq. (\ref{pa}) as
\begin{eqnarray}
&&\gamma_n=\gamma_{n0}+\gamma_{I}, \nonumber\\
&&\Delta\phi=\Delta\phi_0+\Delta\phi_I.
\end{eqnarray}
The noninteracting phases $\gamma_{n0}$ and angle $\Delta\phi_0$ are
\begin{equation}
\begin{aligned}
\gamma_{n0}&=\frac{(2n+1)(1-\sqrt{1-\epsilon^2})T\omega_1}{4\sqrt{1-\epsilon^2}},\\
\Delta\phi_0&=-\int_0^T\left[\frac{\omega_2\epsilon^2\sin^2(\omega_2t)dt}{2\Omega(1-\epsilon\cos(\omega_2t))}
+\frac{\epsilon A_2\sin(\omega_2t)d\Omega}{2\Omega^2}\right],
\end{aligned}\label{gtn}
\end{equation}
and the coupling terms follow
\begin{equation}
\begin{aligned}
\gamma_{I}&=\int^T_0\frac{-\epsilon\omega_1
A_2D^2J[1-\epsilon\cos(\omega_2t)][\epsilon-\cos(\omega_1t)]dt}
{\hbar A_1(1-\epsilon^2)\Omega}\\
&=-\frac J\hbar\Delta\phi_I,\label{pai}
\end{aligned}
\end{equation}
with the definition $D\equiv K/(\sqrt{2\mu_1\mu_2
A_1A_2(1-\epsilon^2)})$ and
$\Omega=A_2\sqrt{1-\epsilon^2-2D^2(1-\epsilon\cos(\omega_1t))(1-\epsilon\cos(\omega_2t))}$.
If we take the limit $D\ll\sqrt{1-\epsilon^2}$, $\Omega\approx
A_2\sqrt{1-\epsilon^2}$ can be treated as a constant, $\gamma_{n0}$
and $\Delta\phi_0$ then satisfy the relation,
\begin{equation}
\gamma_{n0}\approx-\left(n+\frac12\right)\frac{\omega_1\Delta\phi_0}{\omega_2},
\end{equation}
and the coupling Berry phase and Hannay's angle can be approximated
as
\begin{equation}
\begin{aligned}
\gamma_{I}\approx\frac{\epsilon^2 A_2JD^2T\omega_1} {\hbar
A_1(1-\epsilon^2)\sqrt{1-\epsilon^2}},\\
\Delta\phi_I\approx-\frac{\epsilon^2 A_2D^2T\omega_1}{
A_1(1-\epsilon^2)\sqrt{1-\epsilon^2}}.
\end{aligned}
\end{equation}
Considering the elliptic condition ($1-\epsilon^2-2D^2(1+\epsilon)^2
>0$), we perform numerical calculation  for Eq. (\ref{gtn}) and(\ref{pai})
with $\epsilon=\sqrt3/2$, $A_1/A_2=10^{8}$ and $J/\hbar=10^{13}$.
 The results  are presented in Fig. \ref{gamma} and Fig. \ref{theta}.

Fig. \ref{gamma} shows how the Berry phase of the ground state
changes with $K$. The original $0+\delta$ on the horizontal axis
denotes a infinitely small positive number. Because when $K=0$,
there is no interaction between the two subsystems, the range of
integration of the phases and the angle are determined by the period
of $\bm{X}_1$ and $\bm{X}_2$, respectively, rather than their common
period. The relation between Berry phases and Hannay's angle
$\gamma_{n0}=\frac{(2n+1)T\omega_1}{4}\approx-\left(n+\frac12\right)
\frac{\omega_1\Delta\phi_0}{\omega_2}$ agree with the prediction in
Ref. \cite{Berry2}, $\gamma_{n}=(n+\frac12)\pi
=-\left(n+\frac12\right)\Delta\phi$. From the figure, we can find
that the coupling constant $K$ has remarkable  influence on the
Berry phase, and as the coupling increases, the influence becomes
larger. This means that the classical subsystem play a  role of
driving field for the quantum system.  It can also be seen from Fig.
\ref{gamma} that the frequency ratio of the parameters
$\omega_1/\omega_2$ can change  both  Berry phase $\gamma_{I}$ and
$\gamma_{n0}$, since  it determines the common period $T$ of the
parameters $\bm{X}_1$ and $\bm{X}_2$. In contrast, the Hannay's
angle $\Delta\phi_I$ is much smaller than $\Delta\phi_0\approx-\pi
T\omega_2$ as Fig. \ref{theta} shows. Therefore, we conclude that
 not only a classical system can drive a quantum
system to obtain a extra Berry phase, but also the quantum system
can react  back on the classic system and induce a correction to the
Hannay's angle. This is because the quantum system possesses the
same structure of the classical system, and the quantum parallel
transport can be treated as a parallel transport in the effective
classical system. The quantum one-form may also affect the classical
subsystem due to the interaction.

It is worth addressing that this quantum-classical hybrid model is
different from the full quantum coupled quantum generalized harmonic
oscillator model, since they possess different structures. For full
quantum model, the Hamiltonian can be written as
\begin{equation}
\begin{aligned}
\hat{H}=&\frac12[X_1\hat{q}^2+Y_1(\hat{p}\hat{q}+\hat{q}\hat{p})+Z_1\hat{p}^2]+K\hat{q}\hat{Q}\\
&+\frac12[X_2\hat{Q}^2+Y_2(\hat{P}\hat{Q}+\hat{Q}\hat{P})+Z_2\hat{P}^2].
\end{aligned}
\end{equation}
After a canonical transformation for $Q$ and $q$ by
\begin{equation}
\left(
  \begin{array}{l}
    R_1 \\
    R_2 \\
  \end{array}
\right)
=\left(\begin{array}{cc}
    \cos\beta & \sin\beta \\
    -\sin\beta & \cos\beta
  \end{array}\right)
  \left(
  \begin{array}{cc}
    q/\sqrt{Z_1} \\
    Q/\sqrt{Z_2} \\
  \end{array}
\right),
\end{equation}
with
\begin{equation}
\begin{aligned}
&\sin\beta\equiv\left[\frac{\omega_2^2-\omega_1^2+\sqrt{(\omega_1^2-\omega_2^2)^2+4K^2Z_1Z_2}}
{2\sqrt{(\omega_1^2-\omega_2^2)^2+4K^2Z_1Z_2}}\right]^{1/2},\\
&\cos\beta=\left[\frac{\omega_1^2-\omega_2^2+\sqrt{(\omega_1^2-\omega_2^2)^2+4K^2Z_1Z_2}}
{2\sqrt{(\omega_1^2-\omega_2^2)^2+4K^2Z_1Z_2}}\right]^{1/2},
\end{aligned}
\end{equation}
the frequencies of the two oscillators
$\omega_1\equiv(X_1Z_1-Y_1^2)^{1/2}$ and
$\omega_2\equiv(X_2Y_2-Y_2^2)^{1/2}$ change into
\begin{equation}
\begin{aligned}
&\Omega_1\equiv\sqrt\frac{\omega_1^2+\omega_2^2+\sqrt{(\omega^2_1-\omega_2^2)^2+4K^2Z_1Z_2}}{2},\\
&\Omega_2\equiv\sqrt\frac{\omega_1^2+\omega_2^2-\sqrt{(\omega^2_1-\omega_2^2)^2+4K^2Z_1Z_2}}{2}.
\end{aligned}
\end{equation}
The eigenfunctions of $\hat{H}$ can  be written as the product of
the eigenfunctions of two effective harmonic oscillators,
\begin{equation}
\Psi_{mn}(R_1,R_2;\bm{X}_1,\bm{X}_2)=\varphi_{1m}(R_1;\bm{X}_1,\bm{X}_2)
\varphi_{2n}(R_2;\bm{X}_1,\bm{X}_2),\label{fqoeigen}
\end{equation}
where
\begin{equation}
\begin{aligned}
\varphi_{kn}(R_k;\bm{X}_1,\bm{X}_2)=&\left(\frac{\Omega_k}{\hbar}\right)^{1/4}
\chi_n\left(R_k\sqrt\frac{\Omega_k}{\hbar}\right)\\
&\cdot\exp\left(-\frac{iY_1q^2}{2Z_1\hbar}-\frac{iY_2Q^2}{2Z_2\hbar}\right),
\end{aligned}
\end{equation}
and $\chi_{n}$ are the normalized Hermite functions. Inserting Eq.
(\ref{fqoeigen}) into Eq. (\ref{aia}) and Eq. (\ref{anglephase}), we
obtain the one-form and the phase
\begin{equation}
\begin{aligned}
A^B_{mn}=&\frac{-iZ_1}{4}\left[\frac{(2m+1)\cos^2\beta}{\Omega_1}
+\frac{(2n+1)\sin^2\beta}{\Omega_2}\right]d\left(\frac{Y_1}{Z_1}\right)\\
&-\frac{iZ_2}{4}\left[\frac{(2m+1)\sin^2\beta}{\Omega_1}
+\frac{(2n+1)\cos^2\beta}{\Omega_2}\right]d\left(\frac{Y_2}{Z_2}\right),
\end{aligned}
\end{equation}
\begin{equation}
\begin{aligned}
\gamma_{mn}=i&\oint A^B_{mn}\\
=&\oint\left\{\frac{Z_1}{4}\left[\frac{(2m+1)\cos^2\beta}{\Omega_1}
+\frac{(2n+1)\sin^2\beta}{\Omega_2}\right]d\left(\frac{Y_1}{Z_1}\right)\right.\\
&\left.+\frac{Z_2}{4}\left[\frac{(2m+1)\sin^2\beta}{\Omega_1}
+\frac{(2n+1)\cos^2\beta}{\Omega_2}\right]d\left(\frac{Y_2}{Z_2}\right)\right\}.
\end{aligned}\label{fqophase}
\end{equation}
It is easy to find that when there is no interaction between the two
oscillators ($K=0$), the Berry phase in Eq. (\ref{fqophase}) returns
back to a sum of the Berry phases of two generalized harmonic
oscillator\cite{Berry2},
\begin{equation}
\gamma_{mn}=\oint\frac{(2m+1)Z_1}{4\omega_1}d\left(\frac{Y_1}{Z_1}\right)
+\oint\frac{(2n+1)Z_2}{4\omega_2}d\left(\frac{Y_2}{Z_2}\right).
\end{equation}

If $\hat q$ denotes the coordinate of a light particle and $\hat Q$
the coordinate of a  heavy one, we can take Born-Oppenheimer
approximation, treating the heavy particle coordinate  $\hat Q$ as a
parameter for the light particle, the Hamiltonian for the light
particle then takes,
\begin{equation}
\hat{H}_1=\frac12[X_1\hat{q}^2+Y_1(\hat{p}\hat{q}+\hat{q}\hat{p})+Z_1\hat{p}^2]+K\hat{q}Q.
\end{equation}
Its eigenfunctions $\psi_n(q;Q,\bm{X}_1)$ and eigenvalues
$E_n(Q,\bm{X}_1)$ are given by Eq. (\ref{qcoeigen}),
\begin{eqnarray}
&&E_n=\left(n+\frac12\right)\hbar\omega-\frac{Z_1K^2Q^2}{2\omega^2},\nonumber\\
&&\psi_n=\sqrt{\alpha}\chi_n\left(\alpha\left(q+\frac{KZ_1Q}{\omega^2}\right)\right)
\exp\left(\frac{-iY_1q^2}{2Z_1\hbar}\right),\nonumber\\
\end{eqnarray}
with $\omega=\sqrt{X_1Z_1-Y_1^2}$ and
$\alpha=\sqrt{\frac{\omega}{Z_1\hbar}}$. Note that $E_n(Q)$ enters
into the Hamiltonian for the heavy particle as a potential,
Hamiltonian for the heavy particle then takes,
\begin{equation}
H_n^{eff}=\frac12[X_2\hat{Q}^2+Y_2(\hat{P}\hat{Q}+\hat{Q}\hat{P})
+Z_2\hat{P}^2]+E_n(Q,\bm{X}_1).\label{fqoboeff}
\end{equation}
The eigenvalues and eigenfunctions of the effective Hamiltonian can
be calculated straightforwardly,
\begin{equation}
\begin{aligned}
&E^{eff}_{mn}(\bm{X}_1,\bm{X}_2)=\left(m+\frac12\right)\hbar\Omega+\left(n+\frac12\right)\hbar\omega,\\
&\varphi_{m}(Q;\bm{X}_1,\bm{X}_2)=\sqrt{\alpha '}\chi_m(\alpha 'Q)
\exp\left(\frac{-iY_2Q^2}{2Z_2\hbar}\right),
\end{aligned}
\end{equation}
where the effective frequency is defined by
$\Omega=[\frac{(\omega^2X_2-Z_1K^2)Z_2}{\omega^2}-Y_2^2]^{1/2}$
 and $\alpha'=\sqrt{\frac{\Omega}{Z_2\hbar}}$.  $E^{eff}_{mn}(\bm{X}_1,\bm{X}_2)$, $(m,n=1,2,3,...)$ are the
eigenvalues of the total Hamiltonian, the corresponding eigenvectors
are
\begin{equation}
\Psi^{tot}_{mn}(q,Q;\bm{X}_1,\bm{X}_2)\approx\varphi_m(Q;\bm{X}_1,\bm{X}_2)\psi_n(q;Q,\bm{X}_1).
\end{equation}
Therefore, the Berry phase of the total system can be calculated by
Eq. (\ref{aia}) and Eq. (\ref{anglephase}) as
\begin{equation}
\begin{aligned}
\gamma_{mn}=\int&\int dqdQ\oint
\psi_n^{*}(q;Q,\bm{X}_1)\varphi_m^{*}(Q;\bm{X}_1,\bm{X}_2)\\
&\cdot
d_{\bm{X}}[\varphi_m(Q;\bm{X}_1,\bm{X}_2)\psi_n(q;Q,\bm{X}_1)]\\
=\oint&\left\{\left[\frac{(2n+1)Z_1}{4\omega}+\frac{(2m+1)K^2Z_1^2Z_2}{4\omega^4\Omega}\right]d\left(\frac{Y_1}{Z_1}\right)\right.\\
&+\left.\left[\frac{(2m+1)Z_2}{4\Omega}\right]d\left(\frac{Y_2}{Z_2}\right)\right\}.
\end{aligned}\label{fqoophase}
\end{equation}
Interestingly, if we take the Bohr-Sommerfeld quantization rule
$J=(m+1/2)\hbar$ \cite{Berry2,Keller} into account and notice $\oint
d(Y_2/\Omega)=0$, the contribution of $\gamma_{mn}$ to the Berry
phase for the light particle (the first two term in Eq.
(\ref{fqoophase})) exactly matches  the Berry phase for our
quantum-classical hybrid oscillator  in Eq.(\ref{pa}),  and the
contribution to the Berry phase for the heavy particle (the second
two term in Eq. (\ref{fqoophase})) satisfies
$\Delta\phi=-\partial\gamma_{mn}/\partial m$ (see Ref.
\cite{Berry2}), where  the Hannay angle takes Eq.(\ref{pa}). This is
exactly the relation between Berry phase and Hanney's angle.  The
description for quantum and classical system  is different in
physics.  To treat them uniformly, we apply the Weinberg's theory
and express  the quantum subsystem classically. Since the classical
particle is much heavier than the quantum particle, the
Born-Oppenheimer approximation turns to be a good approximation for
this problem, the predictions made in this paper are reasonable.

\section{conclusion}
The Berry phase and Hannay's angle in coupled quantum-classical
hybrid systems have been  studied in this paper.  To calculate
uniformly the Berry phase and Hannay's angle, we introduced a
one-form connection, by which we obtain both of the Berry phase and
the Hannay's angle for the hybrid system. In this sense,  the Berry
phase and Hannay's angle in the quantum and classical subsystem can
be treated uniformly. To illustrate the formalism, we give two
examples. The first example is a spin-half particle coupled to a
classical oscillator. In the second example, we calculated the Berry
phase and the Hannay's angle for two couple oscillators, one of
which is quantum while another is classical.  The effects of
subsystem-subsystem coupling on the phase and angle are given and
discussed. The results show that the classical subsystem provides
the quantum subsystem an large correction  to the Berry phase, while
the quantum subsystem gives the classical Hannay's angle a small
perturbation. These predictions depend on the feature of
quantum-classical hybrid system and their mutual interactions. We
also found that the frequency ratio affects the phases and the
angle, since  it can control the evolution periods  of the quantum
and classical subsystems. Finally,  we have calculated the Berry
phase for a fully  quantum version of the two coupling system.

\ \ \\
This work is supported by NSF of China under grant Nos 61078011 and
10935010, and the Fundamental Research Funds for the Central
Universities.
\\


\begin{thebibliography}{99}
\bibitem{Chruscinski} D. Chru{\' s}ci{\' n}ski, and A.
Jamio{\l}kowski, {\em Geometric Phases in Classical and Quantum
Mechanics} (Birkh{\" a}user, Berlin, 2004).

\bibitem{Berry} M. V. Berry, Roc. R. Soc. A {\bf 392}, 45 (1984).

\bibitem{Berry1} M. V. Berry, ``The quantum phase, five years after"
in {\em Geometric phases in physics}, edited by A. Shapere and F. Wilczek
(World Scientific, Singapore, 1989).

\bibitem{Simon} B. Simon, Phys. Rev. Lett. {\bf 51}, 2167 (1983).

\bibitem{Hannay} J. H. Hannay, J. Phys. A {\bf 18}, 221 (1985).

\bibitem{Arnold} V. I. Arnold, {\em Mathematical methods of classical
mechanics} (Springer-Verlag, Berlin, 1978).

\bibitem{Berry2} M. V. Berry, J. Phys. A {\bf 18}, 15 (1985).

\bibitem{Giavarini} G. Giavarini, E. Gozzi, D. Rohrlich and W. D.
Thacker, Phys. Rev. D {\bf 39}, 3007 (1989).

\bibitem{Jarzynski} C. Jarzynski, Phys. Rev. Lett. {\bf 74}, 1264 (1995).

\bibitem{Pati} A. K. Pati, Ann. Phys. {\bf 270}, 178 (1998).

\bibitem{Heslot} A. Heslot, Phys. Rev. D {\bf 31}, 1341 (1985).

\bibitem{Weinberg} S. Weinberg, Ann. Phys. (N.Y.) {\bf 194}, 336 (1989); S.
Weinberg, Phys. Rev. Lett. {\bf 62}, 485 (1989).

\bibitem{Polchinski} J. Polchinski, Phys. Rev. Lett. {\bf 66}, 397
(1991).

\bibitem{Wu} B. Wu, J. Liu and Q. Niu, Phys. Rev. Lett. {\bf 94}, 140402 (2005).

\bibitem{Zhang} Q. Zhang and B. Wu, Phys. Rev. Lett. {\bf 97}, 190401
(2006).

\bibitem{Stone} M. Stone, Phys. Rev. D {\bf 33}, 1191 (1986).

\bibitem{Gozzi} E. Gozzi and W. D. Thacker, Phys. Rev. D {\bf 35},
2398 (1987).

\bibitem{Shen} J. Shen , X. L. Huang, X. X. Yi, C. F. Wu and C. H.
Oh, Phys. Rev. A. {\bf 82}, 062107 (2010).

\bibitem{Zhan} F. Zhan, Y. Lin and B. Wu, J. Chem. Phys. {\bf 128},
204104 (2008).

\bibitem{Leggett}  A. J. Leggett et al., Rev. Mod. Phys. {\bf 59}, 1 (1987).

\bibitem{Keller} J. B. Keller, Ann. Phys. (NY) {\bf 4}, 180 (1958).

\end{thebibliography}
\end{document}